\documentclass[aps,superscriptaddress,twocolumn,amssymb,amsmath,showpacs]{revtex4}

\usepackage[]{amsfonts}
\usepackage{graphicx}

\newcommand{\be}{\begin{equation}}
\newcommand{\ee}{\end{equation}}
\newcommand{\bea}{\begin{eqnarray}}
\newcommand{\eea}{\end{eqnarray}}
\newcommand{\bh}{\hat{b}}
\newcommand{\bhd}{\hat{b}^\dagger}
\newcommand{\eqname}[1]{\label{eq:#1}}
\newcommand{\eq}[1]{(\ref{eq:#1})}
\newcommand{\omax}{\Omega}
\newcommand{\sinc}{{\rm sinc}}

\begin{document}

\title{Bogoliubov Theory of acoustic Hawking radiation in Bose-Einstein
  Condensates}

\author{A. Recati} \affiliation{Dipartimento di Fisica, Universit\`a di Trento
  and CNR-INFM BEC Center, I-38050 Povo, Trento, Italy}
\affiliation{Physik-Department, Technische Universit\"at M\"unchen, D-85748
  Garching, Germany} 
\author{N. Pavloff} \affiliation{ LPTMS, Universit\'e
  Paris Sud, UMR8626, F-91405 Orsay, France} 
\author{I. Carusotto}
\affiliation{Dipartimento di Fisica, Universit\`a di Trento and CNR-INFM BEC
  Center, I-38050 Povo, Trento, Italy}

\begin{abstract} 
We apply the microscopic Bogoliubov theory of dilute Bose-Einstein condensates to analyze quantum and thermal fluctuations in a flowing atomic condensate in the presence of a sonic horizon. For the simplest case of a step-like horizon, closed-form analytical expressions are found for the spectral distribution of the analog Hawking radiation and for the density correlation function. The peculiar long-distance density correlations that appear as a consequence of the Hawking emission features turns out to be reinforced by a finite initial temperature of the condensate.
The analytical results are in good quantitative agreement with first principle numerical calculations.
\end{abstract}

\maketitle

\section{Introduction}

Thanks to the impressive advances in the cooling and manipulation of ultracold atomic gases, experiments are now able to address novel regimes where the physics of coherent matter wave is strongly affected by zero point quantum fluctuations~\cite{phasefluc,MottBloch,BKTexp,Casimir,Willireview}.
In particular Bose-Einstein condensed (BEC) atomic vapors appear as a versatile
and efficient tool for observing a very fascinating manifestation of quantum fluctuations,
namely Hawking radiation from acoustic black holes, the so-called {\em dumb holes}~\cite{HawkingBEC,LeonhardtBog}.

Hawking radiation is a most celebrated, yet still unobserved prediction of quantum field theory on curved space-times, which consists of the conversion of vacuum fluctuations into observable radiation in the vicinity of a black-hole horizon \cite{Hawking}. 
Elaborating on the mathematical analogy between the propagation of sound waves in inhomogeneous and moving media and the propagation of quantum fields on a curved space–time background, Unruh predicted in 1981 the occurrence of an analog Hawking emission of sound in any system showing a sonic horizon, i.e., an interface between a sub-sonic and a super-sonic region \cite{Unruh}.

Very recently, the experimental realization of a dumb hole-like configuration in a flowing atomic Bose-Einstein condensate was presented in Ref.\cite{TechnionExp}. From a more general standpoint, a dumb-hole configuration for surface waves on a tank of moving water was realized in~\cite{tank}: the reported observation of the conversion of positive-frequency waves into negative frequency ones can be considered as a classical analog of the Hawking effect \cite{Unr95}. 
The observation of a horizon in a microstructured optical fiber was reported in~\cite{Leon-fiber-optic}. 
The possibility of simulating acoustic black holes using a ring-shaped chain of trapped ions was considered in~\cite{cirac-ions}.

So far, much of the theoretical work on the analog Hawking radiation in atomic Bose-Einstein condensates has used some gravitational analogy to characterize the properties of the emission. Our point of view is different and aims at developing a microscopic understanding of analog Hawking radiation starting from the general theory of quantum fluctuations in condensed matter systems and without referring to any gravitational analogy.
A similar approach has been adopted in the few last years by other researchers:  A first step in this direction can be found in Ref. \cite{LeonhardtBog} where a general theory of analog Hawking radiation based on Bogoliubov theory is presented but no detailed analysis of the observable consequences is carried out. This point of view has been pushed further in the very recent papers~\cite{Renaud1,Renaud2}, where the general framework is developed in full detail and numerical results for some observable quantities are also discussed for the case of a smooth sub- to super-sonic flow transition.

In the present paper we report a completely analytical study of Hawking radiation from dumb-holes in atomic Bose-Einstein condensates. Our calculations are based on a direct application of the standard Bogoliubov theory of dilute condensates and no explicit reference is ever made to the gravitational analogy. 
In order to make the problem analytically tractable, we consider a simplest step-like configuration where the transition between the sub-sonic to the super-sonic regions occurs on a very short length scale.
Even if the surface gravity of this configuration is formally infinite, still a thermal-like Hawking emission is found at a temperature fixed by the healing length. Closed analytical formulas for the density correlations are extracted, which are found in excellent agreement with the numerical simulations of~\cite{NJP}. These results extend the analytical understanding of the analog Hawking radiation to the sharp interface limit opposite to the hydrodynamic regime previously considered in the literature. Even if the singularity at the step makes the analogy with gravitational physics {\em strictu sensu} no longer valid, still the remarkable properties of the resulting emission support our choice of calling it {\em analog Hawking radiation}.

The paper is organized as follows. 
In Sec. \ref{model} we present the physical system under consideration and we review the Bogoliubov description of quantum fluctuations in the presence of a sonic horizon.
In Sec. \ref{observables} we derive analytical formulas for the main observable quantities such as the emission spectrum and the density correlation function. These formulas are shown to successfully compare with the numerical results of~\cite{NJP}. The effect of a finite initial temperature is discussed. 
Conclusions are drawn in Sec. \ref{conclusion}.

\section{The physical system and the Bogoliubov description}\label{model}

\begin{figure*}[hptb]
\begin{center}
\includegraphics[width=1.5\columnwidth]{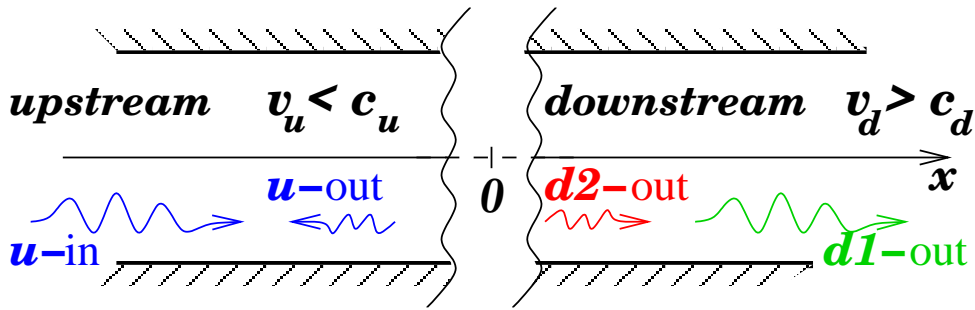}
\caption{(color online) Sketch of the one dimensional dumb hole configuration considered in this paper. A BEC is flowing in the positive $x$ direction. The horizon is located at $x=0$. The flow is uniform in the asymptotic regions far from the horizon, with a velocity which is sub-sonic in the upstream region and supersonic in the downstream one. 
The wiggly arrows illustrate one among the different scattering processes described by Eq.\eq{S}: an incident $u$-in wavepacket is partially reflected onto the $u$-out branch and partially transmitted on the $d1$-out and $d2$-out branches. The chosen labelling of the modes is explained in the text and illustrated in Fig.\ref{fig:disp}).}
\label{sketch}
\end{center}
\end{figure*}

The system we consider is sketched in Fig.\ref{sketch}. A Bose-Einstein condensate is flowing along a one-dimensional atomic waveguide. For simplicity, the transverse trapping is assumed to be tight enough for a
one-dimensional description to be accurate. The flow velocity is directed along the positive-$x$ direction, $v(x)>0$. The density profile $n(x)$ and/or velocity field $v(x)$ of the condensate show a significant spatial modulation in the region around $x=0$.  Far from this interface region, both the density and the flow velocity tend to their up-stream ($x\rightarrow -\infty$) and down-stream ($x\rightarrow +\infty$) asymptotic values $n_{u,d}$ and $v_{u,d}$.  The flow is kept stationary in time by means of a suitable external potential $V_{\rm ext}(x)$ and/or spatial modulation of the atom-atom interaction constant $g(x)$, so that the condensate wavefunction
$\psi_0(x)$ is a solution of the stationary Gross-Pitaevskii equation
\begin{equation}\label{GPE}
H_{GP}\psi_0=\left[-\frac{\hbar^2}{2m}\partial_x^2+V_{\rm ext}(x)+g(x)\,|\psi_0|^2\right]\,\psi_0=\mu\,\psi_0,
\end{equation} 
at a chemical potential $\mu$.
A dumb-hole configuration is realized when the up-stream region is subsonic $v_u<c_u$ and the down-stream region is instead super-sonic $v_d>c_d$. Here, $c_{u,d}=\sqrt{g_{u,d}n_{u,d}/m}$ is the speed of sound and $g_{u,d}=g(x\rightarrow -\infty,+\infty)$ are the asymptotic values of the interaction constant in the $u,d$-regions, respectively. For later purposes, we also introduce the corresponding upstream (downstream) healing length $\xi_{u,d}=\hbar/(m c_{u,d})$, chemical potential $\mu_{u,d}=mc_{u,d}^2$, and condensate wave-vector $k_{u,d}=m\,v_{u,d}/\hbar$.

In this dumb-hole configuration a sound wave emitted in the down-stream region is dragged away by the flow without being able to reach the up-stream region. The down-stream region is thus the sonic analog of the interior of a gravitational black hole, and the transition point where the velocity of the flow is exactly equal to the speed of sound is the analog of the event horizon. We conventionally locate this point at $x=0$. 

Provided the condensate is everywhere dilute $n(x)\,\xi(x)\gg 1$, small fluctuations on top of the condensate can be described by the Bogoliubov theory of dilute condensates~\cite{IvanRev}. In particular, the elementary excitations correspond to the eigenvectors of the Bogoliubov operator:
\begin{equation}
\mathcal{L}=
\left(
\begin{array}{cc}
H_{GP}-\mu+g|\psi_0|^2\ & g\psi_0^2 
\\ 
-g\psi_0^{*2} &
- H_{GP}+\mu-g|\psi_0|^2
\end{array}
\right) \; .
\label{eq:L}
\end{equation}

\subsection{Bogoliubov modes in a homogeneous system}

Within each homogeneous $u$, $d$ region far from the interface, the eigenvectors of
operator (\ref{eq:L}) at an eigenenergy $\hbar \omega$ are plane waves of the form
\begin{equation}\label{bog}
\left(\begin{array}{c} u(x)\\w(x)\end{array}\right)=
\left(\begin{array}{c} \bar{u}(x)\,e^{i k_{u,d} x}\\ \bar{w}(x)\,e^{-i k_{u,d}
x}\end{array}\right)=
e^{i k x}
\left(\begin{array}{r} U_k\,e^{i k_{u,d} x} \vspace{2mm}\\
  W_k\, e^{-i k_{u,d} x} \end{array}\right) \; .
\end{equation}
The wavevector $k$ and frequency $\omega$ satisfy the Bogoliubov dispersion relation 
$\omega=\omega_{\rm B}^{\pm}(k)$ with 
\begin{equation}
\omega_{\rm B}^\pm(k) =
v_{u,d} \, k \pm c_{u,d}\, k \, \sqrt{1+\frac{1}{4}(k\,\xi_{u,d})^2} \; .
\eqname{dispersion}
\end{equation}
As usual in Bogoliubov theory~\cite{IvanRev}, the upper (lower) sign in Eq. \eq{dispersion} refers to the positive
(negative) norm branch, for which $|U_k|^2-|W_k|^2=+1\; (-1)$.

\begin{figure*}[hptb]
\begin{center}
\includegraphics[width=1.2\columnwidth]{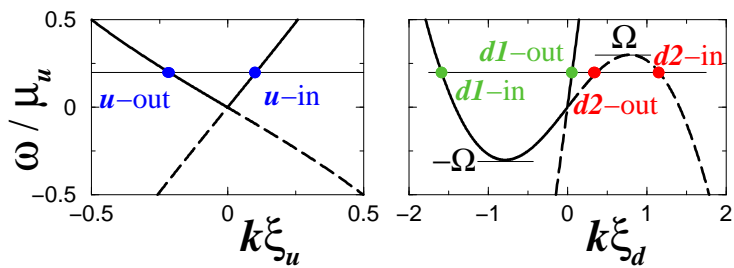}
\caption{(color online) The Bogoliubov mode dispersion relations \eq{dispersion} in the asymptotic upstream (left panel) and in the down-stream region (right panel). 
Solid (dashed) lines correspond to modes with a positive (negative) Bogoliubov norm.}
\label{fig:disp}
\end{center}
\end{figure*}

Let us restrict our attention to the positive frequency modes ($\omega>0$). The negative frequency modes can in fact be obtained from the positive frequency ones by simply reverting the sign of $k$ and exchanging the values of the Bogoliubov coefficients $U$ and $W$; corresponding modes then have opposite norms.

In the up-stream sub-sonic region, for any $\omega>0$ there are two positive norm modes satisfying the Bogoliubov equation $\omega=\omega_{\rm B}^+(k)$ with real wavevectors $k_u^{\rm in}$, $k_u^{\rm out}$ respectively. These modes correspond to propagating plane waves. In the Bogoliubov dispersion shown in Fig.\ref{fig:disp} (left panel), they correspond to the branches indicated as $u$-in and $u$-out. Throughout the whole paper, modes will be labelled as in-going (``in'') if their group velocity $v_g=d\omega/dk$ points toward the horizon and out-going (``out'') in the opposite case. 
In addition to the positive norm modes, a second pair of negative norm of modes exist with complex wavevectors satisfying the Bogoliubov equation $\omega=\omega_{\rm B}^-(k)$: these correspond to exponentially growing or decreasing evanescent waves~\cite{kagan}.

In the down-stream super-sonic region, two positive-norm real solutions of the Bogoliubov dispersion at $k_{d1}^{\rm in}$ and $k_{d1}^{\rm out}$ exist for any $\omega>0$: In the Bogoliubov dispersion of Fig.\ref{fig:disp} (right panel), these wavevectors correspond to the $d1$-in and $d1$-out branches.
On the other hand, two cases have to be distinguished in the negative norm sector, depending on whether $\omega$ is larger or smaller than the critical value $\omax=\omega_{\rm B}^-(K)$ with $K$ defined by
\begin{equation} 
(\xi_d\, K)^2= -2 + \frac{v_d^2}{2\,c_d^2} + \frac{v_d}{2\, c_d}\,
\sqrt{8+\frac{v_d^2}{c_d^2}}\; .  
\end{equation}  
A pair of real solutions $k_{d2}^{\rm in}$ and $k_{d2}^{\rm out}$ to the negative-norm dispersion relation $\omega=\omega_{\rm B}^-(k)$ exist as long as $\omega<\omax$ and correspond to the $d2$-in and $d2$-out branches of Fig.\ref{fig:disp} (right panel). The critical frequency $\omax$ is the maximum of the $d2$-in and $d2$-out branches. For $\omega>\omax$, these real solutions are again replaced by a pair of complex solutions corresponding to evanescent waves.
Of course, the $d2$ branches do not exist if the condensate flow is everywhere sub-sonic.

\subsection{The scattering solution}
\label{sec:scattering}

Far from the interface, the generic eigenfunction of the Bogoliubov operator (\ref{eq:L}) at a frequency $\omega$ is built by superimposing within each $u$, $d$ region all available propagating (that is non-evanescent) plane waves at the given frequency $\omega$:
\begin{eqnarray}
\left(
\begin{array}{c}
\bar{u}(x) \\ \bar{w}(x)
\end{array}
\right)_{u,d}=
\left(\sum_{i \in \textrm{in} } 
\left(
\begin{array}{c}
U_{k_i^{\rm in}} \\ W_{k_i^{\rm in}}
\end{array}
\right)\,
\frac{e^{i k_i^{\rm in} x}}{\sqrt{4 \pi|v_{g,i}^{\rm in}|}} 
\,\beta^{\textrm{in}}_i+\right.\nonumber\\
\left.\sum_{j\in \textrm{out}} 
\left(
\begin{array}{c}
U_{k_j^{\rm out}} \\ W_{k_j^{\rm out}}
\end{array}
\right)\,
\frac{e^{i k_j^{\rm out} x}}{\sqrt{4\pi|v_{g,j}^{\rm out}|}} \,
\beta^{\textrm{out}}_j\right)_{u,d} 
\; .
\label{eq:modes}
\end{eqnarray}
Here, plane waves have been separated into in-going (in) and out-going (out) ones according to the sign of their group velocity. 
Most remarkable among these solutions are the so-called scattering solutions, that describe a plane-wave excitation originating from infinity (either upstream or downstream) on a well defined in-going mode, impinging on the horizon, and then leaving again towards infinity as a superposition of the different out-going branches: only a single in-going $\beta_i^{\rm in}$ amplitude is then non-vanishing, while two (if $\omega>\omax$) or three (if $\omega<\omax$) out-going  $\beta_j^{\rm out}$ components have a finite value, describing reflected and transmitted waves. 

As a specific example, in the case where the in-going $i=d1,d2$ wave originates from the down-stream region, the scattering solution has the form
\begin{multline}
\left(
\begin{array}{c}
\bar{u}(x) \\ \bar{w}(x)
\end{array}
\right)_{d}=
\left(
\begin{array}{c}
U_{k_i^{\rm in}} \\ W_{k_i^{\rm in}}
\end{array}
\right)\,
\frac{e^{i k_i^{\rm in} x}}{\sqrt{4 \pi|v_{g,i}^{\rm in}|}} 
\,\beta^{\textrm{in}}_i+\\
\sum_{j\in \textrm{out}} 
\left(
\begin{array}{c}
U_{k_j^{\rm out}} \\ W_{k_j^{\rm out}}
\end{array}
\right)\,
\frac{e^{i k_j^{\rm out} x}}{\sqrt{4\pi|v_{g,j}^{\rm out}|}} \,
\beta^{\textrm{out}}_j
\; ,\label{scat_d}
\end{multline}
in the down-stream region and the form
\begin{equation}
\left(
\begin{array}{c}
\bar{u}(x) \\ \bar{w}(x)
\end{array}
\right)_{u}=
\left(
\begin{array}{c}
U_{k_u^{\rm out}} \\ W_{k_u^{\rm out}}
\end{array}
\right)\,
\frac{e^{i k_u^{\rm out} x}}{\sqrt{4\pi|v_{g,u}^{\rm out}|}} \,
\beta^{\textrm{out}}_u
\; .\label{scat_u}
\end{equation}
in the up-stream region. 
The two out-going components in Eq. (\ref{scat_d}) correspond to the reflected part in respectively the $d1$ and $d2$ channels; of course, a $d2$ component can only be present if $\omega<\omax$. On the other hand, only a single outgoing transmitted component is present in the upstream region.
Similar expressions can be straightforwardly written in the case where the in-going $u$-in wave originates from the up-stream region.

The expressions (\ref{scat_d}) and (\ref{scat_u}) based on the plane-wave expansion are only valid in the asymptotic regions far from the interface. For any given realization of the dumb-hole configuration the complete
scattering solution for all $x$ can be obtained by solving the full Bogoliubov problem (\ref{eq:L}).
A numerical solution for the case of a smooth interface was presented in~\cite{Renaud2}. An analytical solution for a simplest step-like case will be presented in Sec.\ref{slg}.

In general, the linear relation between the amplitudes $\beta_i^{\rm in}$ and $\beta_j^{\rm out}$ of in- and out-going waves can be written in a compact matricial form~\cite{LeonhardtBog}. 
In the case of a generic flow that remains sub-sonic in both asymptotic regions, as well as for $\omega>\omax$ in a dumb-hole configuration, this relation has the two-by-two form:
\begin{equation}
\left(
\begin{array}{c}
\beta_u^{\textrm{out}}(\omega) \\
\beta_{d1}^{\textrm{out}}(\omega) \\
\end{array}
\right)=
\mathbf{S}(\omega)
\left(
\begin{array}{c}
\beta_u^{\textrm{in}} (\omega)\\
\beta_{d1}^{\textrm{in}} (\omega)\\ 
\end{array}
\right)
\label{eq:S2}
\end{equation}
which only involves positive-norm modes.
On the other hand, the negative-norm $d2$ mode appears as soon as frequencies $\omega<\omax$ are considered. In this case, the matricial relation has the three-by-three form:
\begin{equation}
\left(
\begin{array}{c}
\beta_u^{\textrm{out}}(\omega) \\
\beta_{d1}^{\textrm{out}}(\omega) \\
\beta_{d2}^{\textrm{out}}(\omega)
\end{array}
\right)=
\mathbf{S}(\omega)
\left(
\begin{array}{c}
\beta_u^{\textrm{in}} (\omega)\\
\beta_{d1}^{\textrm{in}} (\omega)\\ 
\beta_{d2}^{\textrm{in}} (\omega)
\end{array}
\right).
\label{eq:S}
\end{equation}

With the chosen normalization of the solution (\ref{eq:modes}), the $\mathbf{S}$ matrix connecting the $\beta$ coefficients in the asymptotic regions is unitary in the Bogoliubov metric $\eta$ inherited by the norm of the
corresponding plane-wave modes
\begin{equation}
\mathbf{S}(\omega)^\dagger \eta \, \mathbf{S}(\omega)=\eta \; .
\label{eq:eta}
\end{equation}
More specifically, $\eta=\textrm{diag}(1,1,-1)$ in the case $\omega<\omax$, whereas $\eta$ is just the standard $2 \times 2$ identity matrix $\eta=\textrm{diag}(1,1)$ for $\omega>\omax$.

Physically, the square moduli $|\mathbf{S}_{ij}|^2$ of the $\mathbf{S}$-matrix elements give the trasmission/reflection coefficients for a $j$-ingoing mode which scatters into an $i$-outgoing mode. The property \eq{eta} ensures total energy conservation. It is interesting to note that a closely related approach was used in the context of quantum evaporation from superfluid $^4$He~\cite{he4}.

As an illustrative example, for each quantum that incides on the system from the $d1$ in-going mode with an energy $\hbar\omega$, one has $|\mathbf{S}_{ud1}|^2$ transmitted quanta in the $u$-out mode, $|\mathbf{S}_{d1d1}|^2$ reflected quanta in the $d1$-out mode, and $|\mathbf{S}_{d2d1}|^2$ quanta reflected in the $d2$-out mode.
While the energy of the transmitted $u$-out and reflected $d1$-out quanta is positive and equal to $\hbar\omega$, the energy of a quantum of the $d2$-out Bogoliubov mode is negative and equal to $-\hbar\omega$~\cite{IvanRev}.
As a result, energy conservation then recovers the unitarity condition $1=|\mathbf{S}_{ud1}|^2+|\mathbf{S}_{d1d1}|^2-|\mathbf{S}_{d2d1}|^2$. 

\subsection{The step like geometry}\label{slg}

All the theory reviewed so far is not limited to a particular realization of dumb-hole and can be applied to generic configurations as long as the density and flow velocity tend to constant values in the asymptotic regions far from the interface.
A numeric solution for a smooth interface was indeed reported in~\cite{Renaud2}.

In the present paper, we restrict our attention to a model step-like structure with
$V_{\rm ext}(x)= V_{\rm ext}^u\Theta(-x)+V_{\rm ext}^d\Theta(x)$ and $g(x)=g_u\Theta(-x)+g_d\Theta(x)$ for which analytical formulas for the components of the $\mathbf{S}$-matrix can be found in closed form. As usual, $\Theta$ is here the Heavyside step-function. 
Provided one imposes~\cite{NJP}
\begin{equation}\label{constantdensity}
V_{\rm ext}^u+\mu_u = V_{\rm ext}^d + \mu_d \; ,
\end{equation}
the time-independent Gross-Pitaevskii equation (\ref{GPE}) has a solution in the plane-wave form
\begin{equation}
\psi_0(x)=\sqrt{n_0}\exp{(ik_0 x)}
\end{equation}
that describes a flow with uniform density $n_{u,d}=n_0$ and velocity $v_{u,d}=v_0=\hbar k_0/m$. If one chooses $c_d<v_0<c_u$, a dumb-hole configuration is obtained. The value of the potential jump $V_{\rm ext}^u-V_{\rm ext}^d$ is then fixed by the condition (\ref{constantdensity}).  This model configuration has the remarkable advantage of allowing for analytical insight without having to solve any differential equation. 
In a future publication we will present the analysis of other realistic dumb-hole configurations \cite{future}.

In the step like geometry, the expressions (\ref{scat_d}) and (\ref{scat_u}) for the scattering solution can be extended to the whole $x>0$ and $x<0$ regions simply by including evanescent modes as well~\cite{kagan}. 
A complete scattering solution is then formed by the superposition of five two dimensional column vectors of the form (\ref{bog}): one of these corresponds to the incoming ($i=u,d1,d2$) mode and the four others are outgoing waves (two reflected and two transmitted). Depending on the value of $\omega$, some of the outgoing waves may be evanescent: in this case, between the two conjugate complex wavevectors that solve the Bogoliubov dispersion \eq{dispersion}, only the wavevector value giving an exponentially decaying wave at large distance from the horizon has to be considered. 

Across the step-structure at $x=0$ the column vector eigenfunction of ${\cal L}$ and its first derivative have to be continous. 
This provides the four matching conditions that are necessary to determine the amplitudes of the four outgoing waves as a function of the amplitude $\beta_i^{\rm in}$ of the in-going $i$ one. The coefficients of the non evanescent outgoing waves correspond to the matrix elements $\mathbf{S}_{ji}$ with $j=u$, $d1$, $d2$. As usual, the $d2$ element only exists for $\omega>\omax$. 
Note that correct inclusion of the evanescent waves is here crucial to be able to fulfill the four matching conditions.

\subsection{Transmission and reflection}

\begin{figure}[ptb]
\begin{center}
\includegraphics[height=6.cm] {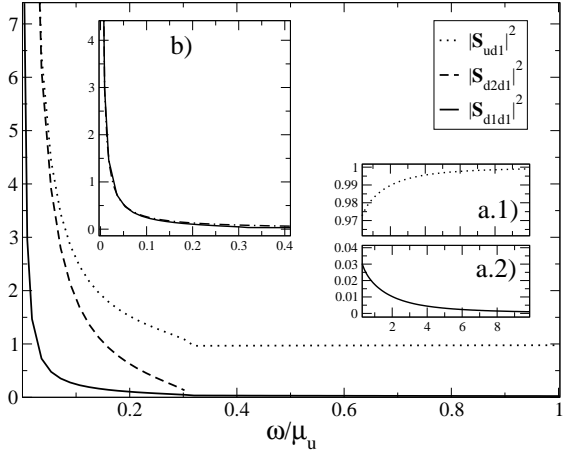}
\caption{Transmission and reflection coefficients for a $d1$ in-going mode on a
  step-like dumb hole configuration.  $|{\mathbf S}_{u d1}|^2$ (dotted curve)
  corresponds to the transmission on the $u$-out mode; $|{\mathbf S}_{d1 d1}|^2$
  (solid curve) and $|{\mathbf S}_{d2 d1}|^2$ (dashed curve) correspond to the
  reflection on the $d1$-out and $d2$-out modes respectively. 
The parameters $v_0/c_u=0.75$, $v_0/c_d=1.5$ used here are the same as in Ref. \cite{NJP}. \
Above $\omax$ ($\omax/\mu_u=0.316$ for the chosen set of parameters) one recovers (insets a.1) and a.2) the usual case of a single transmitted ($u$-out) and a single reflected ($d1$-out) waves.
  respectevely.  Inset b) displays a comparison of the exact $|\mathbf{S}_{d1 d1}|^2$ coefficient (solid line) with its low energy approximation (\ref{sd1d1}) (dot-dashed line).}
\label{fig:TR}
\end{center}
\end{figure}

The typical behaviour of the scattering coefficients for a $d1$ ingoing wave is plotted in Fig.\ref{fig:TR} as a function of the frequency $\omega$. 
For $\omega>\omax$, the transmitted $u$-out wave and the reflected $d1$-out wave are only present. As usually expected in wave mechanics, the transmission (reflection) coefficient increases (decreases) as a function of $\omega$, see the insets (a1-a2). Energy conservation imposes that the sum of these two coefficients is always equal to unity.

For $\omega<\omax$ also the (negative-norm) reflected mode $d2$-out is involved in the dynamics. Remarkably, all three scattering coefficients diverge as $1/\omega$ in the low $\omega$ limit. Nonetheless, energy conservation is ensured by the $\eta$-unitarity of the $\mathbf{S}$-matrix, which now imposes $|S_{ud1}|^2+|S_{d1d1}|^2-|S_{d2d1}|^2=1$. 

Straightforward algebraic manipulations lead to a simple analytical formula for the reflection coefficient in the low-$\omega$ limit,
\begin{equation}
|\mathbf{S}_{d1d1}|^2\simeq 
\frac{c_u}{c_d}\, \frac{(c_u-v_0)(c_u-c_d)}{(c_u+v_0)(c_u+c_d)}
\left(\frac{v_0^2}{c_u^2}-\frac{c_d^2}{c_u^2}\right)^{3/2}
\frac{\mu_u}{2 \hbar \omega}:  \label{sd1d1}
\end{equation}
the excellent agreement between this low energy approximation and the exact Bogoliubov
result is apparent in Fig. \ref{fig:TR}(b).

\begin{figure}[ptb]
\begin{center}
\includegraphics[width=0.95\columnwidth]{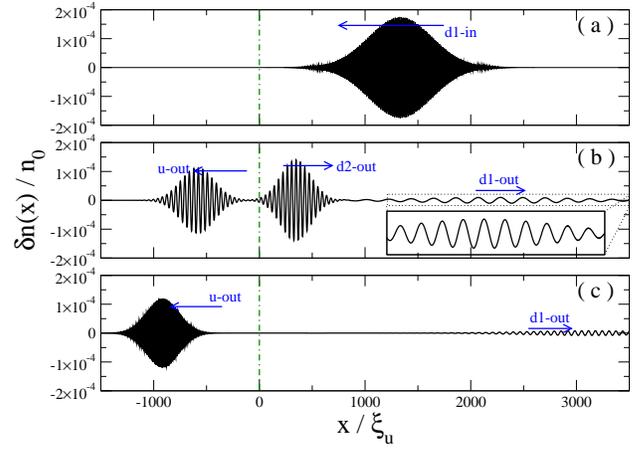}
\caption{A $d1$ ingoing wave packet is incident on a step-like sonic horizon located at $x=0$ (upper panel). When the central  frequency omega of the wave packet is smaller that Omega, a single transmitted $u$ wavepacket and two $d1,d2$ reflected wave packets are visible (middle panel). A single $u$ trasmitted and a single $d1$ reflected wave packet are instead visible for $\omega>\omax$ (lower panel). Same system parameters as in Fig.\ref{fig:TR}. Wave-packet carrier wave-vector $q\,\xi_u=-1.2$ (upper and middle panels), $q\,\xi_u=-1.35$ (lower panel).
}
\label{fig:packets}
\end{center}
\end{figure}

The wave mechanics encoded in the $\mathbf{S}$-matrix is physically illustrated in Fig.\ref{fig:packets} where we show the result of a numerical solution of the time-dependent Gross-Pitaevskii equation that describes the scattering of a incident $d1$ in-going wave packet (upper panel) onto the black hole horizon. 
For a low initial energy, the in-going wave packet splits into a $u$-out transmitted one (central panel),
and a pair of $d1$, $d2$ reflected ones: Among these, the bigger and slower wavepacket corresponds to the negative-norm $d2$-out mode, while the other one corresponds to the $d1$-out mode. 
As expected, for an in-going wavepacket energy above $\Omega$ (lower (c) panel), only the $u$ transmitted and the $d1$ reflected wavepackets are visible.
A related phenomenology was experimentally observed for surface waves in a water tank in~\cite{tank} and theoretically discussed in~\cite{cirac-ions} for a chain of trapped ions. In~\cite{zapata}, this negative norm reflected mode was interpreted in terms of a bosonic analog of Andreev reflection.

\subsection{Quantization of the modes}

In the presence of a macroscopically occupied condensate, the full Bose field operator can be expanded as
\begin{equation}
\hat\psi(x)=\psi_0(x)+\hat{\delta\psi}(x)\,.
\end{equation}
The coherent condensate is described by a classical field $\psi_0$ that evolves according to the Gross-Pitaevskii equation. Quantum fluctuations are described by the operator term $\hat{\delta\psi}$ whose dynamics is ruled by the Bogoliubov operator \eq{L}.

A most favourable basis to our purposes consists of the scattering solutions at a given frequency $\omega$ that we have discussed in Sec.\ref{sec:scattering}. Within each $\omega$ sub-space, the three scattering modes corresponding to the $I=u,d1,d2$ in-going modes form an ortho-normal basis (of course, one has to restrict to $I=u,d1$ for $\omega>\omax$) in the Bogoliubov $\eta$-metric,  
\begin{eqnarray}
\int\!\!\!&dx&
({u_{I\omega}(x)^*}u_{J\omega'}(x)-{w_{I\omega}^*(x)}w_{J\omega'}(x))\nonumber\\ && = s_I\delta_{IJ}\delta(\omega-\omega')\\
\int\!\!\!&dx&
({u_{I\omega}(x)}v_{J\omega'}(x)-{v_{I\omega}(x)}u_{J\omega'}(x))=0 \; .
\end{eqnarray}
The $s_I$ coefficient gives the sign of the Bogoliubov norm of the mode: in our case, one has $s_I=+1$ for $I=u,d1$ and $s_I=-1$ for $I=d2$. Spatial integration is over the whole space.

The final form of the quantum field is obtained by the standard Bogoliubov quantization prescription~\cite{IvanRev}: the amplitudes corresponding to positive-norm modes become destruction operators while amplitudes of negative-norm modes become creation operator,
\begin{multline}
\hat{\delta\psi}(x)= \int_0^\infty\!\!\!\!\!
d\omega\!\!\!\sum_{I=u,d1}\left[u_{I\omega}(x)\,\hat a_I(\omega)+w_{I\omega}^*(x)
\, \hat a_I(\omega)^\dagger\right]  \\
+\int_0^\infty\!\!\! d\omega \left[u_{d2,\omega}(x)\,\hat a^\dagger_{d2}(\omega)+w_{d2,\omega}^*(x)\,\hat a_{d2}(\omega)\right].
\label{eq:flucfield}
\end{multline}
Expectation values of physical observables (e.g. the density correlation function) are then straightforwardly evaluated by imposing the suitable boundary conditions on the expectation values of products of in-going operators, $\hat{a}_I$ and $\hat{a}^\dagger_I$.

If one is interested in correlation functions involving the out-going modes only, a reformulation of \eq{flucfield} in the input-output language~\cite{input-output} can be used, as proposed in~\cite{LeonhardtBog}. In our case, the input-output relations consist of a linear relation connecting the operators of the out-going modes $\bh_{u,d1,d2}$ to the in-going $\hat{a}_{u,d1,d2}$ ones via the $\mathbf{S}$ matrix \eq{S} introduced in the previous subsection:
\begin{equation}
\left(
\begin{array}{c}
\bh_u(\omega) \\
\bh_{d1}(\omega) \\
\bhd_{d2}(\omega)
\end{array}
\right)=
\mathbf{S}(\omega)
\left(
\begin{array}{c}
\hat{a}_u(\omega)\\
\hat{a}_{d1}(\omega)\\ 
\hat{a}^\dagger_{d2}(\omega)
\end{array}
\right).
\label{eq:Sq}
\end{equation}
As a consequence of their negative Bogoliubov norm, the $d2$ modes appear in both the left- and the right-hand side of \eq{Sq} as creation operators rather than destruction ones. As we shall see in Sec.\ref{sec:spectrum}, this simple fact is the mathematical origin of the Hawking emission in our formalism. From a quantum optical perspective, Hawking emission can then be interpreted as parametric down-conversion of Bogoliubov sound waves by the horizon.

\section{observables}\label{observables}

The expression \eq{flucfield} for the quantum field in the Bogoliubov approximation and the input-output relation \eq{Sq} are the starting point for the calculation of physical observables of the system: in particular, our attention will be focussed on two most remarkable ones, namely the spectral distribution of the Hawking emission and the long-distance behaviour of the correlation function of density fluctuations.  

\subsection{Emission spectrum}
\label{sec:spectrum}
In the gravitational context, the only observable quantity is the Hawking radiation outside the
black hole. One of the most remarkable feature is that
this radiation is thermal at a temperature univocally determined
by the surface gravity of the black hole.

In our condensed matter context, the Hawking emission from the dumb hole corresponds to the phonons that are emitted into the upstream region on the $u$-out branch. 
Assuming that no correlation between the different in-going modes exist, the emission spectrum, that is the number of phonons emitted per unit time and per unit bandwidth into the $u$-out branch is straightforwardly calculated from the input-output relation \eq{Sq}:
\begin{multline}
\frac{dI^{\rm out}_u}{dt\,d\omega}=\langle \hat{b}_{u}^\dagger(\omega)\, \hat{b}_{u}(\omega) \rangle= \\ =|\mathbf{S}_{uu}|^2\,I_u^{\rm in}+|\mathbf{S}_{ud1}|^2\,I_{d1}^{\rm in}+|\mathbf{S}_{ud2}|^2\,(I_{d2}^{\rm in}+1)\,.
\end{multline}
For a system initially at zero-temperature, all the $I_{u,d1,d2}^{\rm in}=\langle \hat{a}_{u,d1,d2}^\dagger(\omega)\, \hat{a}_{u,d1,d2}(\omega) \rangle=0$, yet a finite Hawking emission exists as a consequence of the $+1$ term arising from the $\langle \hat{a}_{d2}(\omega)\, \hat{a}_{d2}^\dagger(\omega) \rangle$ expectation value that encodes quantum fluctuations. 

Restricting our attention to the low-frequency part of the spectrum, an analytical form can be found for the emission spectrum:
\begin{multline}\label{sud2} 
\left.\frac{dI^{\rm out}_u}{dt\,d\omega}\right|_{T=0}=|\mathbf{S}_{ud2}(\omega)|^2\simeq \\
\frac{(c_u-v_0)}{(c_u+v_0)}\frac{c_u^2}{(c_u^2-c_d^2)}
\left(\frac{v_0^2}{c_u^2}-\frac{c_d^2}{c_u^2}\right)^{3/2}
\frac{2\mu_u}{\hbar \omega} \;. 
\end{multline}
It is remarkable to note that this formula still gives the typical the $1/\omega$ thermal behavior of Hawking radiation even though one is not allowed to use the gravitational analogy: in the present step-like case, the surface gravity is in fact formally infinite. 
The effective temperature of the emission, i.e. the coefficient of the $1/\omega$ term, is here determined by the microscopic physics of the condensate that fixes the cut-off frequency $\omax$ and the chemical potential $\mu_{u,d}$.

\subsection{Density correlations}

As first remarked in Ref.\cite{Correlation}, the density correlation function appears to be the most promising tool for identifying Hawking radiation: in particular, this quantity was at the heart of the numerical observation of~\cite{NJP}.
An example of such a calculation is reproduced in Fig.\ref{fig:G2newMC}. 
The dark regions correspond to antibunching, and the bright ones to bunching. The strongest feature in this figure is the dark stripe along the $x=x'$ line: as its origin is well known and corresponds to the antibunching due to the repulsive inter-atomic interaction~\cite{Nar99}, we will not discus it further. In the present paper, we will rather focus our attention onto the other features of the correlation plot, labeled as $u-d2$, $u-d1$ and $d1-d2$, that encode the information on the Hawking emission.

Within Bogoliubov theory, the density correlation function in the stationary state
\begin{eqnarray}\label{g2}
G^{(2)}(x,x') =
\frac{\langle\hat{\psi}^\dagger(x)\hat{\psi}^\dagger(x')
\hat{\psi}(x')\hat{\psi}(x)\rangle}{\langle\hat{\psi}^\dagger(x)\hat{\psi}(x)\rangle \langle\hat{\psi}^\dagger(x')
\hat{\psi}(x')\rangle}-1\; .
\end{eqnarray}
can be expanded in its $\omega$ components as:
\begin{equation}\label{inte-omega}
G^{(2)}(x,x')=\int_\mathbb{R^+} G^{(2)}(\omega,x,x')\, d\omega \; .
\end{equation}
Each $\omega$ component has to be evaluated imposing suitable boundary conditions on the expectation values of in-going operator averages.
In the following of the discussion, it will reveal useful to separate the zero temperature contribution $G^{(2)}_{0}$ from the thermal one $G^{(2)}_{\rm th}$ 
\begin{equation}
G^{(2)}(\omega,x,x')=G^{(2)}_{0}(\omega,x,x')+G^{(2)}_{\rm th}(\omega,x,x')\; .
\end{equation}
The former term only involves the zero-point fluctuations of the in-going modes
\begin{multline}
n_0G^{(2)}_{0}(\omega,x,x')=
\frac{1}{4\pi}\Big[ \sum_{I=u,d1}w_{I\omega}^*(x)\, r_{I\omega}(x') \\ 
+u_{d2\omega}^*(x)\,r_{d2\omega}(x')+\rm{c.c.}\Big] \; ,
\end{multline}
while the latter includes the initial populations $\langle a_I^\dagger(\omega)\,a_I(\omega)\rangle$:
\begin{multline}
n_0G^{(2)}_{\rm th}(\omega,x,x')=\frac{1}{4\pi}
\sum_{I=u,d1,d2}\left[r_{I\omega}^*(x)\,r_{I\omega}(x')+\rm{c.c.}\right]\times \\
\times\langle a_I^\dagger(\omega)\,a_I(\omega)\rangle \; .
\end{multline}
In both these formulas, we have set $r_I=u_I+w_I$.

\subsection{Zero Temperature density correlations}

\subsubsection{In-Out correlation}
Let us first consider the in-out zero-temperature correlations between the inside ($x'>0$) and the outside ($x<0$) regions. For $x$ and $x'$ far from the interface, it is legitimate to only keep the terms with a stationary phase and neglect the fast oscillating ones. Making use of the $\eta$-unitarity of the scattering matrix and setting $R_k=U_k+W_k$, the simple expression
\begin{multline}
n_0G^{(2)}_{0}(\omega,x,x')= \frac{1}{4\,\pi} \sum_{l=d1,d2}
\Big\{
\frac{R_{k^{\rm out}_u}(\omega)R_{k^{\rm out}_l}(\omega)}
{\sqrt{|v^{\rm out}_{g,u}\,v^{\rm out}_{g,l}|}}
\times \\
e^{i [k^{\rm out}_u(\omega) x-k^{\rm out}_l(\omega) x']}
\mathbf{S}^*_{u d2}(\omega)\,\mathbf{S}_{l d2}(\omega)+\rm{c.c.}\Big\},
\label{eq:G2relevant}
\end{multline}
is obtained for the long-distance zero-temperature correlation function.
As all the $\mathbf{S}$-matrix elements appearing in Eq.\eq{G2relevant} involve the $d2$ mode, it is apparent that a non-zero long-distance correlation is only possible in a dumb-hole configuration: in terms of the input-output picture of \eq{Sq}, the two terms in \eq{G2relevant} originate from the quantum fluctuations of the ingoing $d2$ mode that are scattered in either the $u$ or the $l=d1$, $d2$ modes. 

\begin{figure}[ptb]
\begin{center}
\includegraphics[height=6.3cm] {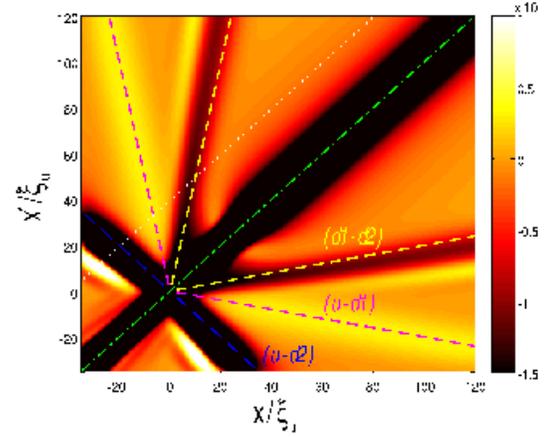}
\caption{(color online) Plot of the rescaled density correlation
$n_0\,\xi_u\,[G^{(2)}_0(x,x')-1]$ for
  an initial zero temperature. The data are the same as the one presented in
  \cite{NJP}, but the color scale has been modified in order to make the
  positive $u$-$d1$ correlation clearly visible. This signal was overlooked in
  the original paper given its relative weakness.
  The dashed lines identify the different feature of the correlation function
  as discussed in the text.  The white dotted line corresponds to the cut
  $x'-x=39.5\; \xi_u$ used in Figs. \ref{cutud2} and \ref{fig:MC}.}
\label{fig:G2newMC}
\end{center}
\end{figure}

As the integrals in (\ref{inte-omega}) are dominated by the small $\omega$ part of the integral, simple analytical expressions can be worked out by extrapolating the low-$\omega$ asymptotic form of the $\mathbf{S}$-matrix elements 
\begin{equation}
\mathbf{S}^*_{u d2}(\omega)\,\mathbf{S}_{l d2}(\omega)\simeq \frac{A_l\mu_u}{\hbar\omega}
\end{equation}
to the whole spectrum and then cutting the integral at $\omax$. 
In this hydrodynamic approximation, $k^{\rm out}_j\simeq\omega/V_j$ ($j=u$, $d1$, $d2$) and the velocities $V_j$ of the outgoing modes are $V_u=v_0-c_u<0$, $V_{d1}=v_0+c_d>0$, $V_{d2}=v_0-c_d>0$.
As typical of Bogoliubov theory~\cite{IvanRev}, the $R_k$ coefficients giving the density weight of the Bogoliubov modes scale as $\sqrt{\omega}$. 
The in-out correlation function then takes the simple form
\begin{equation}
n_0G^{(2)}_{0}(x,x')\simeq\!\!\!\sum_{l=d1,d2}\frac{A_l\,c_u^2\,\omax\,
  \sinc\left(\omax(\frac{x}{V_u}-\frac{x'}{V_l})\right)}{4\pi
  |V_u V_l|\sqrt{c_u c_d}} \; ,
\label{eq:G2app}
\end{equation} 
where $\sinc(x)=\sin x /x$ and the $A_l$ coefficient has the following explicit form  
\begin{equation}\label{eq:Al} 
A_{d1\;(d2)}=\sqrt{\frac{c_u}{c_d}}
\frac{(c_u-v_0)}{(c_u+v_0)}
\left(\frac{v_0^2}{c_u^2}-\frac{c_d^2}{c_u^2}\right)^{3/2}
\frac{c_u}{c_d+(-) c_u} \; 
\end{equation}
in terms of the microscopic parameters of the system. 

From the explicit expression \eq{G2app}, it is immediate to see that the density correlation is maximum along the pair ($l=d1,d2$) of straight lines
\begin{equation}
\frac{x}{V_u}=\frac{x'}{V_{l}}
\label{eq:straight}
\end{equation}
that exit in opposite directions $x<0$, $x'>0$ from the horizon position $x=x'=0$. The absence of lateral shift from the horizon position $x=x'=0$ is a consequence of the reality of the $A_l$ coefficients. As a consequence of the inequality $c_u>v_0>c_d$, the sign of the $d2$ contribution is positive, while the one of the $d1$ contribution is negative. The $d2$ contribution is always stronger than the $d1$ one by a significant factor 
\begin{equation}
\frac{|A_{d2}/V_{d2}|}{|A_{d1}/V_{d1}|}=\frac{(c_u-c_d)(v_0-c_d)}{(c_d+c_u)(v_0+c_d)}
\end{equation}

The $d2$ contribution corresponds to the Hawking signature anticipated in \cite{Correlation} and numerically observed in~\cite{NJP}. Once the horizon is formed, the quantum fluctuation of the incoming negative-norm $d2$ mode start being converted into correlated pairs that emerge from the horizon in the $u$ and $d2$ modes. The two phonons are simultaneously created at the horizon and then propagate away at speeds $V_{u}=v_0-c_u<0$ and $V_{d2}=v_0-c_d>0$. At time $t$ after their emission, they are therefore located at $x=V_{u} t<0$ and $x'=V_{d2} t>0$, which explains the correlation between the density fluctuations at points verifying $x/V_{u}=x'/V_{d2}$.

This simple argument can be straightforwardly extended to explain the geometry of the other $d1$ feature. This feature went overlooked in~\cite{NJP} because of its weak intensity, but was mentioned in~\cite{Renaud2}. When the numerical data of~\cite{NJP} are plotted with a suitable color scale as done in Fig.\ref{fig:G2newMC}, the $u$-$d1$ correlation is perfectly visible as a positive correlation tongue.

\begin{figure}[ptb]
\begin{center}
\includegraphics[width=0.9\columnwidth]{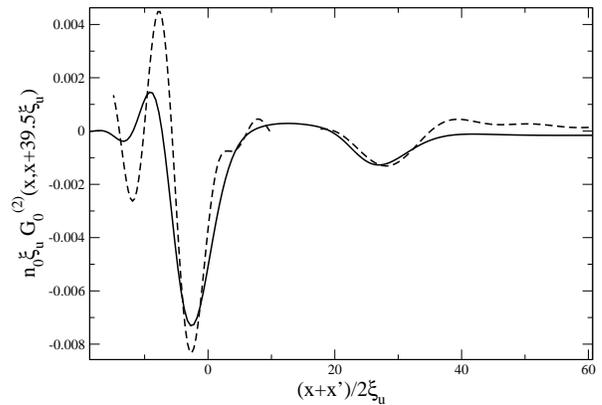}
\caption{Cut of the zero-temperature density correlation function $G_0^{(2)}$ along the $x'-x=39.5\,\xi_u$ line. 
Solid line: numerical result from~\cite{NJP}. Dashed lines: prediction of Eqs.\eq{G2relevant} and \eq{G2relevantII} for respectively the $u-d2$ and the $d1-d2$ contributions. On this scale the $u-d1$ contribution is hardly visible.}
\label{fig:MC}
\end{center}
\end{figure}

A quantitative comparison of the prediction \eq{G2relevant} with the numerical results is illustrated in Fig.\ref{fig:MC} where cuts of the density correlation function along the line $x'-x=39.5\,\xi_u$ (marked by a white dashed line in Fig.\ref{fig:G2newMC}) are shown using the same set of parameters as in Ref.\cite{NJP}.
In spite of the approximations made, the agreement is remarkable.

\begin{figure}[ptb]
\begin{center}
\includegraphics[height=5.5cm] {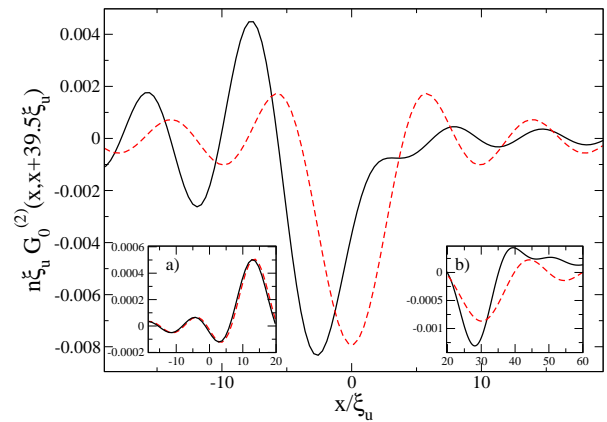}
\caption{(Color online) Cut of the density correlation function $G_0^{(2)}$ along the $x'-x=39.5 \,\xi_u$ line. Black solid lines are the prediction of \eq{G2relevant} and \eq{G2relevantII}. Red dashed lines are the hydrodynamical approximations \eq{G2app} and \eq{G2IIapp}.
The main panel shows the $u-d2$ feature in the in-out region. Inset (a) shows the $u-d1$ feature in the in-out region. Inset (b) shows the $d1-d2$ feature in the in-in region.}
\label{cutud2}
\end{center}
\end{figure}

The accurateness of hydrodynamical approximation \eq{G2app} is finally validated in Fig.\ref{cutud2}.
As shown in the main panel, the agreement for the $l=d2$ feature is quite good even though the hydrodynamic approximation is not fully able to reproduce the asymmetry of the peak nor its slightly back-shifted position with respect to the straight line \eq{straight}. Both these facts can be traced back to the quick deviation from a linear dispersion that is visible in Fig.\ref{fig:disp} (right panel) for the $d2$ Bogoliubov branch.
This interpretation is confirmed by the excellent agreement which is instead found for the $l=d1$ contribution (inset (a) of Fig.\ref{cutud2}): the discrepancy of the dispersions of both the $u$ and the $d1$ modes from the hydrodynamic approximation is in fact very small in the whole frequency range up to $\omega=\omax$.

\subsubsection{In-In correlation}

The in-in correlation function for a pair of points $x,x'$ both located inside $x,x'>0$ the dumb hole can be calculated along the same lines.
In this case, a single contribution appears that originates from the correlation between the $d1$ and the $d2$ modes,
\begin{multline}
n_0G^{(2)}_{0}(\omega,x,x')= \frac{1}{4\,\pi}
\left(
\frac{R_{k^{\rm out}_{d1}}(\omega)R_{k^{\rm out}_{d2}}(\omega)}
{\sqrt{|v^{\rm out}_{g,d1}v^{\rm out}_{g,d2}|}}
\times\right. \\
\times \left. e^{i (k^{\rm out}_{d1}(\omega) x-k^{\rm out}_{d2}(\omega) x')}\,
\mathbf{S}^*_{d1 d2}(\omega)\,
\mathbf{S}_{d2 d2}(\omega)+\rm{c.c.}\right) \\
+(x\leftrightarrow x').
\label{eq:G2relevantII}
\end{multline}
Even though it did not appear in the analytical calculations of~\cite{Correlation} based on the gravitational analogy, this contribution was observed and discussed in the numerical paper~\cite{NJP}.

The same hydrodynamical approximation that led to \eq{G2app} can be used to obtain an approximate expression to this $d1-d2$ feature as well. 
For small $\omega$, the ${\mathbf S}$-matrix product can be expanded as 
\begin{equation}
\mathbf{S}^*_{d1 d2}(\omega)\,\mathbf{S}_{d2 d2}(\omega)\simeq\frac{B\mu_u}{\hbar\omega}
\end{equation}
with
\begin{equation}\label{B}
B=-\frac{c_u}{2c_d}
\frac{(c_u-v_0)}{(c_u+v_0)}\left(\frac{v_0^2}{c_u^2}-\frac{c_d^2}{c_u^2}\right)^{3/2}
\; .
\end{equation}
Extrapolating this expression to the whole $\omega$ spectrum and cutting the integral at $\omax$, one immediately gets to the expression
\begin{multline}
n_0{G^{(2)}_{0}(x,x')}\simeq
\frac{B\,\omax\,c_u^2\,\sinc\left(\omax(\frac{x}{V_{d1}}-
\frac{x'}{V_{d2}})\right)}{4\pi V_{d1}V_{d2}\, c_d}+ \\
+(x\leftrightarrow x')\,.
\label{eq:G2IIapp}
\end{multline}

\subsection{Finite Temperature density correlations}

One of the major issues in view of an experimental observation of the analog Hawking radiation is the role of temperature. To this purpose, a naive, but extremely constraining requirement is often imposed that the temperature of the system should be (much) smaller than the Hawking temperature of the emission.
The numerical simulations of \cite{NJP} have instead shown that correlations are robust with respect to temperature, which supports once more their promise in view of an actual experiment.
In the present section we extend the theory of correlations developed in the previous section to the case of a condensate at a finite initial temperature.

In particular, we focus our attention on the same {\em dynamical} situation already considered in~\cite{NJP}: The condensate initially has a uniform density $n_0$, flow velocity $v_0$ and sound velocity $c_u$ and is at thermal equilibrium in the moving frame at a temperature $T$ \cite{footnote}. The horizon is then (adiabatically) switched on by ramping down the scattering length in the downstream region.

Within each semi-infinite uniform region, the population of each $k$ mode is adiabatically preserved during the formation of the dumb hole. The initial population $\langle a_I^\dagger a_I\rangle$ of each $I=u,d1,d2$ mode at wave vector $k_I$ is given by the thermal population $n(\Omega_I) = [\exp(\hbar \Omega_I/k_B T)-1]^{-1}$ corresponding to its frequency
\begin{equation}
\hbar \, \Omega_I=
\sqrt{\frac{\hbar^2 k_I^2}{2m}
\left(\frac{\hbar^2 k_I^2}{2m}+2mc_u^2\right)} \; .
\label{eq:bigomega}
\end{equation}
in the comoving frame before the creation of the horizon. 
For each $\omega$, the wavevector $k_I$ of the corresponding in-going mode has to be determined by inverting the relation $\omega=\omega_{\rm B}^{\pm}(k_{I})$.

The density correlation function at a finite temperature can then be written as a sum of different terms,
\begin{multline}
G^{(2)}(\omega,x,x')= G^{(2)}_{0}(\omega,x,x')
\left[1+n(\Omega_{d1})+n(\Omega_{d2})\right] \\
+ F_{SH}(\omega,x,x')+F_{NSH}(\omega,x,x').
\label{eq:finiteT}
\end{multline}
Here, $n(\omega)= [\exp(\hbar \omega/k_B T)-1]^{-1}$ is the thermal law at a temperature $T$.

The first term in the r.h.s. of \eq{finiteT} is straightforwardly interpreted as a thermal enhancement of the zero-point signal described in the previous subsection, i.e. a {\em stimulated} Hawking emission. 
Other scattering processes peculiar to the $T\neq 0$ case are responsible for the other contributions, which can be separated in a $u-d2$ term due to the presence of a sonic hole,
\begin{multline}
n_0 F_{SH}(\omega,x,x')=
\frac{R_{k_u^{\rm out}}R_{k_{d2}^{\rm out}}
e^{i (k_u^{\rm out} x-k_{d2}^{\rm out} x')}}
{4\pi\sqrt{|v_{g,u}^{\rm out}v_{g,d2}^{\rm out}|}}\times \\
\mathbf{S}^*_{u u}\,\mathbf{S}_{d2 u}(n(\Omega_u)-n(\Omega_{d1}))+\rm{c.c.} \; ,
\label{eq:G2TH}
\end{multline}
and a $u-d1$ term that instead  persists even in absence of the sonic hole:
\begin{multline}
n_0F_{NSH}(\omega,x,x')=
\frac{R_{k_u^{\rm out}}R_{k_{d1}^{\rm out}}}
{4\pi\sqrt{|v_{g,u}^{\rm out}v_{g,d1}^{\rm out}|}}
e^{i (k_u^{\rm out} x-k_{d1}^{\rm out} x')}\times\\
\mathbf{S}^*_{u u}\,
\mathbf{S}_{d1 u} (n(\Omega_u)-n(\Omega_{d1}))+\rm{c.c.} \; .
\label{eq:NSH}
\end{multline}

In the low-$\omega$ limit the $\mathbf{S}$-matrix element products involved in the new terms \eq{G2TH} and \eq{NSH} tend to finite values
\begin{equation}
\mathbf{S}^*_{u u}(\omega)\,\mathbf{S}_{d1(d2) u}(\omega)\simeq-\sqrt{\frac{c_u}{c_d}}\frac{c_u-v_0}{c_u+v_0}\frac{c_d+(-)v_0}{c_u+v_0},
\end{equation}
while the thermal population is proportional to $1/\omega$,
\begin{equation}
n(\Omega_u)-n(\Omega_{d1})\simeq \frac{v_0+c_u}{c_u}\frac{k_BT}{\hbar\omega}.
\end{equation}
Combining these two facts, it is immediate to see that in the hydrodynamic approximation the contributions of \eq{G2TH} and \eq{NSH} to respectively the $u-d1$ and $u-d2$ features then the same form as the zero-temperature one \eq{G2app}.

Summing up all the terms, the only effect of a finite initial temperature on the $d1-d2$ feature is a bosonic stimulation factor $1+n(\Omega_{d1})+n(\Omega_{d2})$. In particular, this feature is not affected by the $F_{SH}$ and $F_{NSH}$ terms.

The $u-d2$ feature is affected by this multiplicative factor as well as by the additional term $F_{SH}$: remarkably, the zero-point \eq{G2app} and thermal \eq{G2TH} contributions have opposite signs. Within the hydrodynamic approximation, the ratio of their contributions to the peak correlation signal has the simple expression
\begin{equation}
-\frac{c_u(v_0-c_d)(c_u-c_d)}{(v_0^2-c_d^2)^{3/2}}\,\frac{k_B T}{\mu_u}\;.
\end{equation}
As usual, the zero-point contribution dominates at low temperature while it is overwhelmed by the thermal one at high temperature. For the parameters used for the finite temperature numerical simulations in Ref.~\cite{NJP}, the $T=0$ was still the most relevant one.

The effect of the thermal occupation on the $u-d1$ feature is even more dramatic.
Also in this case, the (amplified) zero-point positive feature due to the $l=d1$ term in \eq{G2app} and the thermal contribution \eq{NSH} have opposite signs. As the involved scattering process is not limited to the $\omega<\omax$ window, the term \eq{NSH} gets contributions from high frequencies and is for this reason generally larger in magnitude than \eq{G2TH}.
For this reason, the zero-point contribution only dominates at very low temperatures and is quickly overwhelmed by the thermal one: this effect is clearly visible when comparing the $T=0$ numerical simulations shown in Fig.\ref{fig:MC} with the finite temperature ones shown in Fig.6(a) of Ref.~\cite{NJP}.

When the flow remains everywhere subsonic, all long-distance features of the density correlation function disappear but for the positive $u-d1$ tongue at finite temperature. This fact was apparent in Figs.4(a) and 6(b) of Ref.\cite{NJP} and corresponds within the present formalism to the $F_{NSH}$ term. Its physical origin can be traced back to the finite reflectivity of the interface region even in the absence of an horizon and the different initial occupation of the $u$ and $d1$ incident modes. 

Before concluding, it is important to stress that a different configuration was considered in~\cite{Renaud2}, where the dumb hole was assumed to be in thermal equilibrium in the comoving frame {\em after} completion of the horizon formation process. Of course, this configuration can be still described by Eq.\eq{finiteT}, but the initial populations $n(\Omega_{d1,d2})$ have to be computed using in \eq{bigomega} the $c_d$ instead of $c_u$ for the downstream region.

\section{Conclusion}\label{conclusion}

In this work we have made use of the Bogoliubov theory of dilute Bose-Einstein condensates to study the analog Hawking emission that is emitted by an acoustic black holes. Our framework does not rely on any gravitational analogy and is able to provide closed-form analytical formulas for the emission spectrum and the density correlation function of a simplest step-like configuration. Although the surface gravity is in this case formally infinite, the low-energy part of the emission spectrum still shows a thermal character, but the temperature is found to depend on the microscopic properties of the condensate. The extension of our theory to the case of a non-vanishing initial temperature is discussed. We have shown that this general framework is able to quantitatively reproduce the results of first principle numerical calculations~\cite{NJP}, and to provide a physical understanding of the various features that appear in the density-density correlation function. Generalization of the present theory to more general dumb-hole configurations is in progress~\cite{future}.

\begin{acknowledgments}
We are grateful to F. Sols, I. Zapata, R. Parentani, and W. Zwerger for fruitful discussions. Continuous exchanges with R. Balbinot and A Fabbri are warmly acknowledged. This work was supported by the IFRAF Institute, by
Grant ANR-08-BLAN-0165-01, by Ministero dell’Istruzione, dell’Universit\`a e
della Ricerca (MIUR) and by the EuroQUAM FerMix program.
\end{acknowledgments}


\begin{thebibliography}{99}
\bibitem{phasefluc} D. S. Petrov, D. M. Gangardt, and G. V. Shlyapnikov, J.
Phys. IV France {\bf 116}, 5 (2004).

\bibitem{Casimir}
J. N. Fuchs, A. Recati, W. Zwerger, Phys. Rev. A {\bf 75}, 043615 (2007);
J. Schiefele and C. Henkel, J. Phys. A: Math. Theor. {\bf 42}, 045401 (2009).

\bibitem{BKTexp}
Z. Hadzibabic, P. Kruger, M. Cheneau, B. Battelier, and J. Dalibard, Nature {\bf 441}, 1118 (2006).

\bibitem{MottBloch}
M. Greiner, M. O. Mandel, T. Esslinger, T. H\"ansch, and I. Bloch, Nature 
{\bf 415}, 39 (2002).

\bibitem{Willireview}
I. Bloch, J. Dalibard, W. Zwerger, Rev. Mod. Phys. {\bf 80}, 885 (2008).

\bibitem{HawkingBEC} L. J. Garay, J. R. Anglin, J. I. Cirac and P. Zoller,
  Phys. Rev. Lett. {\bf 85} 4643 (2000); C. Barcel\'o, S. Liberati and
  M. Visser, Int. J. Mod. Phys. A {\bf 18} 3735 (2003); C. Barcel\'o,
  S. Liberati and M. Visser, Phys. Rev. A {\bf 68} 053613 (2003);
  S. Giovanazzi, C. Farrell, T. Kiss and U. Leonhardt, Phys. Rev. A {\bf 70},
  063602 (2004); C. Barcel\'o, S. Liberati and M. Visser, Living
  Rev. Rel. {\bf 8} 12 (2005); R. Sch\"utzhold, Phys. Rev. Lett. {\bf 97} 190405
  (2006); S. Wuester and C. M. Savage, Phys. Rev. A {\bf 76} 013608 (2007).

\bibitem{LeonhardtBog} 
U. Leonhardt, T. Kiss, P. Ohberg, J. Opt. B: Quantum  Semiclass. Opt. 
{\bf 5}, S42 (2003).


\bibitem{Hawking} S. W. Hawking, Nature {\bf 248}, 30 (1974);
  Commun. Math. Phys. {\bf 43} 199 (1975). 

\bibitem{Unruh}
W. G. Unruh, Phys. Rev. Lett. {\bf 46}, 1351 (1981).

\bibitem{tank} G. Rousseaux, C. Mathis, P. Maissa, T. G. Philbin,
  U. Leonhardt, New J. Phys. {\bf 10}, 053015 (2008).

\bibitem{Unr95} W. G. Unruh, Phys. Rev. D {\bf 51}, 2827 (1995).

\bibitem{Leon-fiber-optic} 
Thomas G. Philbin, Chris Kuklewicz, Scott Robertson, Stephen Hill, Friedrich König, Ulf Leonhardt1, Science {\bf 319}, 1367 (2008).

\bibitem{cirac-ions} B. Horstmann, B. Reznik, S. Fagnocchi and J. I. Cirac,
  arXiv:0904.4801.

\bibitem{TechnionExp}
O. Lahav, A. Itah, A. Blumkin, C. Gordon, J. Steinhauer, arXiv:0906.1337.  

\bibitem{Correlation} R. Balbinot, A. Fabbri, S. Fagnocchi, A. Recati, and
  I. Carusotto, Phys. Rev. A {\bf 78}, 021603 (2008).

\bibitem{NJP} I. Carusotto, S. Fagnocchi, A. Recati, R. Balbinot, and
  A. Fabbri, New J. Phys. {\bf 10}, 103001 (2008).

\bibitem{Renaud1} J. Macher and R. Parentani, arXiv:0903.2224

\bibitem{Renaud2} J. Macher and R. Parentani, arXiv:0905.3634.

\bibitem{IvanRev} Yvan Castin, in {\sl Coherent atomic matter waves}, Lecture
  Notes of Les Houches Summer School, edited by R. Kaiser, C. Westbrook, and
  F. David, EDP Sciences and Springer-Verlag (2001).

\bibitem{kagan} Yu. Kagan, D. L. Kovrizhin, and L. A. Maksimov, Phys. Rev.
Lett. {\bf 90}, 130402 (2003).

\bibitem{he4} F. Dalfovo, M. Guilleumas, A. Lastri, L. Pitaevskii and
  S. Stringari, J. Phys.: Condens. Matter {\bf 9}, L369 (1997).

\bibitem{future}
N. Pavloff, A. Recati and I. Carusotto, in preparation.


\bibitem{input-output} D.F. Walls and G.J. Milburn, \textit{Quantum Optics} (Springer-Verlag, Berlin, 1994).

\bibitem{zapata}  I. Zapata and F. Sols, Phys. Rev. Lett. {\bf 102}, 180405 (2009).

\bibitem{Nar99} M. Naraschewski and R. J. Glauber, Phys. Rev. A {\bf 59}, 4595
  (1999).  


\bibitem{footnote} This corresponds to the typical experimental situation where an external potential (or a jump in scattering length) is swept at constant velocity through a condensate initialy at rest~\cite{defect}.

\bibitem{defect} 
C. Raman, M. K\"ohl, R. Onofrio, D. S. Durfee, C. E. Kuklewicz, Z. Hadzibabic, and W. Ketterle, Phys. Rev. Lett. {\bf 83}, 2502 (1999);
R. Onofrio, C. Raman, J. M. Vogels, J. R. Abo-Shaeer, A. P. Chikkatur, and W. Ketterle, Phys. Rev. Lett. {\bf 85}, 2228 (2000); 
P. Engels and C. Atherton, Phys. Rev. Lett. {\bf 99}, 160405 (2007).
\end{thebibliography}
\end{document}